\documentclass[twocolumn,amssymb,nobibnotes,aps,prbstab, reprint,letterpaper,pre/printnumbers,amsmath,
superscriptaddress,longbibliography,
showpacs,floatfix]{revtex4-1}
\usepackage{dcolumn}
\usepackage{bm}
\usepackage{graphics}
\usepackage{float}
\usepackage{epsfig}
\usepackage{multirow}
\usepackage{amsmath}
\usepackage{multirow}
\usepackage{tabularx}
\usepackage{mathrsfs}
\usepackage{tabularx}
\usepackage{tabulary}
\usepackage[dvipsnames]{xcolor}
\usepackage[pass,paperwidth=8.5in,paperheight=11in]{geometry}

\usepackage[linktocpage=true]{hyperref}
\usepackage{hyperref}
\usepackage{comment}
\usepackage{soul}
\usepackage{notes2bib}
\hypersetup{
  pdfnewwindow=true, 
  colorlinks=true,
  linkcolor=blue, 
  anchorcolor=blue,
  citecolor=blue, 
  filecolor=blue,
  menucolor=blue, 
  urlcolor=blue}

\usepackage[normalem]{ulem}
\usepackage{xcolor}
\usepackage{orcidlink}

\begin{document}

\title{Probing the Nature of Interstitial Anionic Electrons in 2D Electride Ca$_2$N via Landau-Level Spectroscopy}

\author{Arjyama Bordoloi\,\orcidlink{0009-0006-2760-3866}}
\affiliation{Department of Mechanical Engineering, University of Rochester, Rochester, New York 14627, USA}

\author{Daniel Kaplan}
\affiliation{Center for Materials Theory, Department of Physics and Astronomy, Rutgers University, Piscataway, New Jersey 08854-8019, USA}

\author{Sobhit Singh\,\orcidlink{0000-0002-5292-4235}}
\email{s.singh@rochester.edu}
\affiliation{Department of Mechanical Engineering, University of Rochester, Rochester, New York 14627, USA}
\affiliation{Materials Science Program, University of Rochester, Rochester, New York 14627, USA}
\affiliation{Center for Coherence and Quantum Optics, University of Rochester, Rochester, NY, 14627, USA}


\begin{abstract}
We investigate the magnetic-field response of interstitial anionic electrons (IAEs) in two-dimensional electrides, 
using monolayer Ca$_2$N as a prototypical system. 
By computing the Landau-level (LL) spectrum of the electride bands forming the Fermi surface, we find a linear LL evolution with magnetic field that closely resembles the behavior of a nearly-free 2D electron gas (2DEG). 
The extracted cyclotron effective mass and Landé g-factor deviate moderately from their free-electron values, 
indicating that the IAEs retain a remarkably free-electron-like character. 
Furthermore, the energy dispersion of the electride bands remains insensitive to the choice
of exchange-correlation functional (LDA vs.~PBEsol), indicating that local exchange and correlation effects have minimal influence on the IAEs.
Overall, our findings provide fundamental insight into the quantum nature of electrides and open new avenues for exploring magnetic confinement, correlation effects, and emergent quantum phenomena in low-dimensional interstitial electronic systems.
\end{abstract}

\maketitle

\section{Introduction}
Unlike conventional solids -- such as molecular, covalent, and ionic compounds with localized valence electrons, or metals with itinerant electrons that freely traverse the crystal lattice -- electrides are a unique class of materials where valence electrons reside in the interstitial voids of crystal lattice, often referred as interstitial anionic electrons (IAEs)~\cite{Ellaboudy_JACS_1983,Singh_Nature_1993,Sushko_PhysRevLett_2003,Matsuishi_Science_2003}. In recent years, electrides have attracted significant research interest due to their unique physical and chemical properties, which are primarily governed by the degree of localization and the specific crystallographic sites occupied by the IAEs~\cite{zurek_JACS_2011, racioppi_Annual_Review_of_Mat_Research_2025}. 
Among the different classes of electrides, categorized on the basis of dimensionality and spatial distribution of their IAEs, layered or two-dimensional (2D) electrides have garnered particular attention. These materials exhibit a wide range of novel properties, including high electron mobility~\cite{Lee_Nature_2013}, efficient charge transport~\cite{Kim_Chem_Sci_2015}, magnetism arising from anionic free electrons~\cite{Lee_Nat_com_2020,Lee_npj_quantum_2021,Zhou_PhysRevB_2020}, as well as topological~\cite{Hirayama_PhysRevX_2018} and superconducting characteristics~\cite{Zeng_PhysRevB_2018,Wang_PhysRevB_2024}.

A major breakthrough in the field of low-dimensional electrides was the experimental realization of Ca$_2$N, a layered electride~\cite{Lee_Nature_2013}, which has also been successfully exfoliated down to its monolayer form~\cite{Druffel_JACS_2016}. 
The presence of IAEs in the interlayer gaps, unbound to any nucleus, offers an ideal channel for charge transport. This nucleus-free electron transport mechanism endows Ca$_2$N with an exceptionally high electron mobility of 189\,cm$^2$V$^{-1}$s$^{-1}$~\cite{Zeng_PhysRevB_2018}. Additionally, the material exhibits a low work function, attributed to its highly delocalized, high-density IAEs\,(1.39\,$\times$\,10$^{22}$cm$^{-3}$)~\cite{Druffel_JACS_2016}. 
These features make Ca$_2$N a strong candidate for applications in inorganic catalysis and next-generation electronic devices.

Despite significant advances in exploring the application-oriented properties of  Ca$_2$N, the fundamental nature of the IAEs in Ca$_2$N, and electrides in general,  remains ambiguous and only partially understood. For instance, ab initio calculations of the dielectric response in Ca$_2$N by Cudazzo \textit{et al.} revealed an intrinsically negative in-plane dispersion of the anionic plasmons, indicating that the behavior of these IAEs deviates significantly from that of a homogeneous electron gas~\cite{Cudazzo_PhysRevB_2017}. In contrast, weak electron-phonon coupling observed in Ca$_2$N monolayers suggests that the spatial distribution of IAEs may resemble a two-dimensional electron gas in free space~\cite{Zeng_PhysRevB_2018}. Some other studies, however, propose a highly correlated nature for these electrons~\cite{Novoselov_Journal_of_phy_chem_C_2021}. Taken together, these varying interpretations suggest that the precise nature of IAEs warrants further investigation.

One promising approach to understanding the nature of IAEs is to investigate their response to external magnetic fields. Quantization of IAEs into Landau levels~(LLs) under strong magnetic fields could provide valuable insights into their effective mass, spin-orbit coupling effects, and exchange interactions, as well as the effective $g$-factors that can be extracted from LL splitting. These parameters could serve as useful benchmarks for comparing the nature of IAEs with conventional two-dimensional electron gases (2DEGs). Importantly, since the effective g-factor can be measured experimentally, this approach provides a practical and feasible method to probe the intrinsic properties of IAEs.  This direction, however, remains largely unexplored in the existing literature. 


Building on this insight, in this work we investigate the magnetic field response of IAEs in Ca$_2$N monolayers. Our results show that the LLs of these IAEs exhibit a linear dispersion with the applied magnetic field, with associated $g$ factors ranging from approximately 1 to 2. While the effective mass of these electrons differs significantly from the free electron mass, the LL dispersion indicates that IAEs in Ca$_2$N monolayers behave similarly to a nearly free electron gas. These findings are expected to be applicable to other 2D electrides as well, since electride states are minimally affected by atomic spin-orbit coupling due to their weak interaction with nuclear potentials, and generally exhibit highly dispersive parabolic bands in most cases. Since the electride states typically cross the Fermi level, 2D electrides could provide an ideal platform for realizing textbook Landau quantization theories in real materials, which have traditionally been studied only in idealized 2DEGs.

\section{Computational details}
\label{comp details}

First-principles density functional theory (DFT) calculations~\cite{HK_dft_1964, KS_dft_1965} were performed on the free-standing Ca$_2$N monolayer using the projector augmented wave (PAW)~\cite{Blochl94} method, as implemented in the Vienna Ab initio Simulation Package (VASP) \cite{Kresse96a, Kresse96b, KressePAW}. The Brillouin zone was sampled using a $\Gamma$-centered $k$-mesh of $12 \times 12 \times 1$, and a plane-wave kinetic energy cutoff of 650 eV was employed. Structural relaxation was carried out until the residual Hellmann–Feynman forces on each atom were less than $10^{-3}$ eV/Å, and the total electronic energy was converged to within $10^{-7}$ eV. The PAW pseudopotentials included eight valence electrons for Ca (3p$^6$4s$^2$) and five for N (2s$^2$2p$^3$). A vacuum thickness of more than 15 Å was introduced along the out-of-plane direction to eliminate spurious interactions between periodic images of the monolayers. To evaluate the influence of exchange–correlation effects on the interstitial anionic electrons, we computed the electronic band structure using both the local-density approximation (LDA)~\cite{LDA_CA} and the generalized gradient approximation as parametrized by Perdew, Burke, and Ernzerhof for solids (PBEsol)~\cite{PBEsol}. As the electride bands showed negligible sensitivity to the choice of exchange-correlation functional, all subsequent calculations were carried out using the PBEsol functional. The electronic band structure was also computed both with and without including spin-orbit coupling (SOC); however, SOC was found to have a minimal impact on the electronic properties of the monolayer.

\begin{figure*}
\centering
\includegraphics [width=1\textwidth]{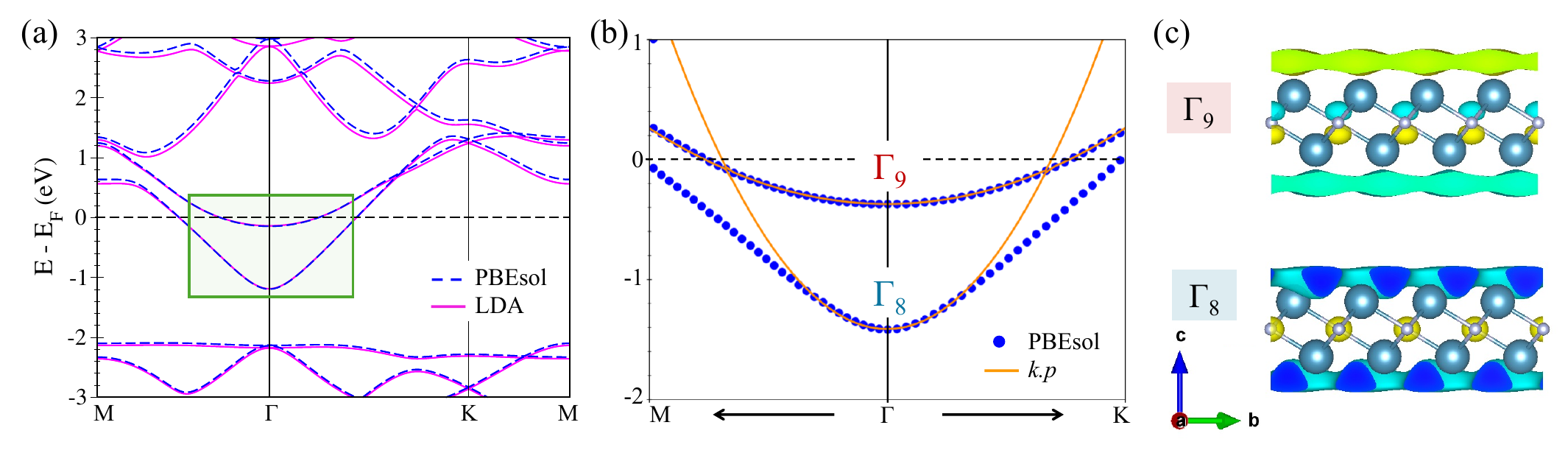}
\caption{(a)Electronic band structure of the Ca$_2$N monolayer computed along high-symmetry directions of the Brillouin zone using PBEsol and LDA functionals. (b)Band structure fitted using the $\mathbf{k} \cdot \mathbf{p}$ model developed for the Ca$_2$N monolayer. (c)Real-space visualization of the electride band wavefunctions using VESTA.
}
\label{fig:band structure}
\end{figure*}

To investigate the Landau level spectrum of the Ca$_2$N monolayer, we constructed an effective $\mathbf{k \cdot p}$ Hamiltonian as a power series expansion in $\mathbf{k}$, using the method of invariants as implemented in the Python packages DFT2KP~\cite{DFT2KP_1,DFT2KP_2} and QSYMM~\cite{QSYM_Varjas_2018}. This was based on the electronic band structure data obtained from QUANTUM ESPRESSO~\cite{QE-2009,QE-2017}. The basis set for the $\mathbf{k \cdot p}$ model was chosen to be the two doubly degenerate electride states with $\Gamma_8$ and $\Gamma_9$ symmetry, which dominate the low-energy regime near the Fermi level, as highlighted in Fig.~\ref{fig:band structure}.

\begin{figure}[!!t]
\centering
\includegraphics [width=9cm]{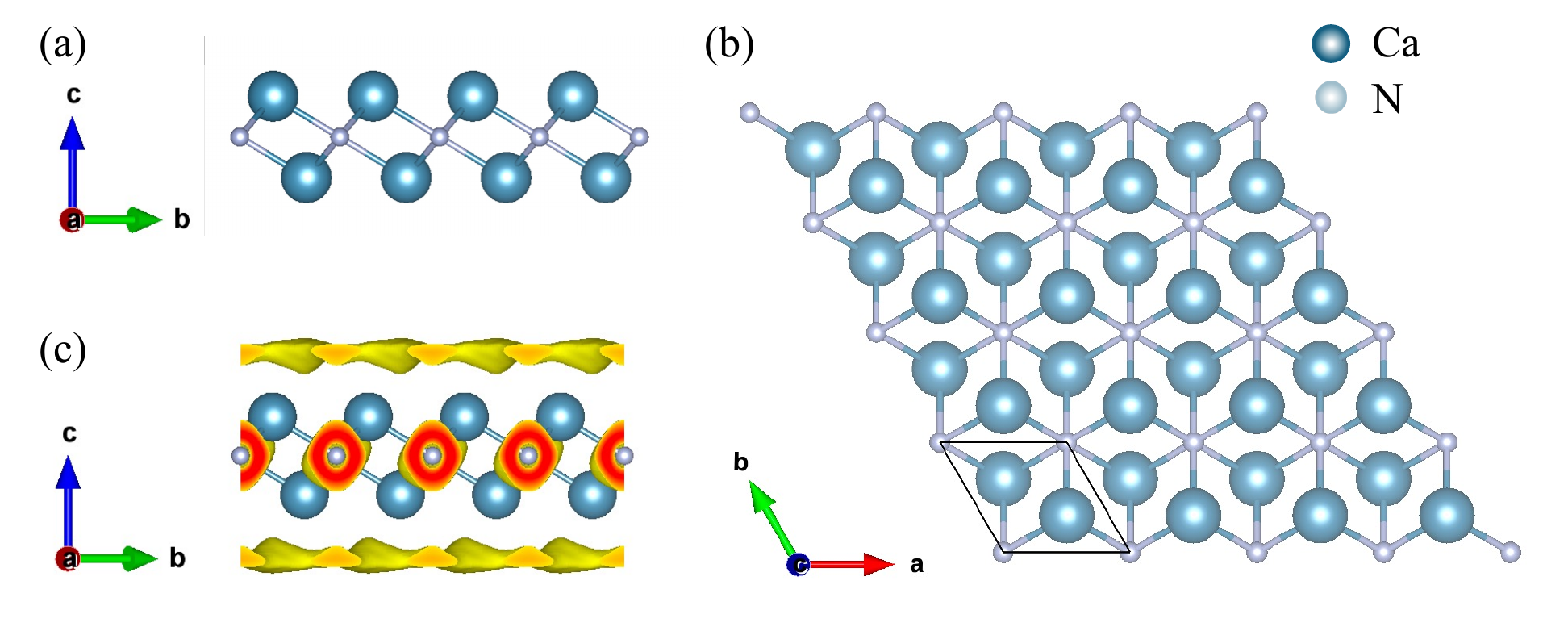}
\caption{
(a) Side view and (b) top view of the crystal structure of Ca$_2$N monolayer.
(c) Electron localization function (ELF) of Ca$_2$N monolayer, visualized using VESTA with an isosurface value of 0.65.}
\label{fig:structure}
\end{figure}

\section{Results and discussions}
\subsection{Crystal structure }

Figure~\ref{fig:structure} shows the crystal structure of the Ca$_2$N monolayer. It adopts an anti-CdCl$_2$ type structure and belongs to the layer group $p\bar{3}m1$ (no.\,\#\,72), as depicted in Figure~\ref{fig:structure}(a)-(b). Consequently, the monolayer belongs to the point group D\textsubscript{3d}, which is characterized by three symmetry generators: $C_{3z}$ (a $2\pi/3$ rotation about the $z$-axis), $C_{2x}$ (a $\pi$ rotation about the $x$-axis), and inversion symmetry $C_i$. These generators are employed in constructing the $\mathbf{k \cdot p}$ Hamiltonian for the monolayer, as discussed in the subsequent sections. The optimized lattice parameters are found to be $a = b = 3.564 \, \text{\AA}$, which are in good agreement with previously reported values~\cite{Druffel_JACS_2016}. We also compute the electron localization function (ELF) of the monolayer, shown as an isosurface at a value of 0.65 in Figure~\ref{fig:structure}(c). The pronounced localized electron cloud near the surface of the monolayer corresponds to the IAEs, confirming the electride nature of Ca$_2$N and consistent with earlier studies~\cite{Druffel_JACS_2016,Zeng_PhysRevB_2018}.

\subsection{Electronic band structure}

Figure~\ref{fig:band structure}(a) represents the electronic band structure of the Ca$_2$N monolayer, computed along high-symmetry directions of the Brillouin zone using both the LDA and PBEsol exchange-correlation (XC) functionals. The dashed line represents the Fermi level. As evident from Figure~\ref{fig:band structure}(a), the Fermi surface is primarily composed of two bands: a highly dispersive band with $\Gamma_8$ symmetry and a less dispersive band with $\Gamma_9$ symmetry. To understand the origin of these two bands, we plotted the corresponding real-space wavefunction profiles at the $\Gamma$-point, as shown in Figure~\ref{fig:band structure}(c). The spatial distribution of these wavefunctions strongly resembles the localized electron cloud observed in the ELF plot, confirming that the electride state arises from a superposition of these two partially occupied bands. 
Consequently, both $\Gamma_8$ and $\Gamma_9$ bands are identified as electride states and a basis set comprising these two states is used to construct the effective $\mathbf{k} \cdot \mathbf{p}$ Hamiltonian, as explained in the following section.

Notably, the dispersion of the electride bands, i.e., the $\Gamma_8$ and $\Gamma_9$ bands remains unchanged upon switching the XC functional from LDA to PBEsol, whereas other bands farther from the Fermi level exhibit slight shifts. This insensitivity of the electride states to the choice of XC functional suggests that the IAEs are largely unaffected by local exchange and Coulomb interaction effects, consistent with the Landau level spectrum results presented in section~\ref{LL}.

We also examined the influence of SOC by calculating the electronic band structure both with and without SOC. For the Ca$_2$N monolayer, the band structures in the two cases show virtually no discernible difference. This insensitivity is primarily due to the weak intrinsic SOC of calcium and nitrogen atoms. To further probe the role of SOC in the electride states, we artificially enhanced the SOC strength within VASP. Remarkably, the electride bands remained completely unaffected by this artificial scaling, whereas other bands displayed noticeable SOC-induced modifications. This behavior can be attributed to the fact that the IAEs are localized near the surface of the monolayer and are thus not under influence of any nuclear potential. As a result, they are largely immune to atomic SOC effects. This finding suggests that such robustness against SOC should be a general feature of electride states in other 2D electrides as well.

\subsection{Landau quantization in Ca$_2$N monolayers}
\label{LL}
To investigate the impact of a magnetic field on the IAEs, we first construct an effective $\mathbf{k \cdot p}$ Hamiltonian for the low-energy bands near the Fermi level. As explained in the previous section, we select the $\Gamma_8$ and $\Gamma_9$ bands as the basis set and use the electronic band structure data computed with SOC to build the effective Hamiltonian. To obtain the $\mathbf{k \cdot p}$ expansion around the $\Gamma$-point, we employ the matrix representations of the symmetry generators of the $D_{3d}$ point group, namely $C_{3z}$, $C_{2x}$, and inversion $C_i$ along with time-reversal symmetry $\mathcal{T}$, expressed in terms of the basis functions corresponding to the $\Gamma_8$ and $\Gamma_9$ states. This leads to a symmetry-constrained expansion of the well-known Kane model around the \( \Gamma \)-point of the Brillouin zone. Since both $\Gamma_8$ and $\Gamma_9$ bands are doubly degenerate, the resulting $\mathbf{k \cdot p}$ Hamiltonian is a $4 \times 4$ matrix, expanded up to second order in $\mathbf{k}$. The first-order and second-order coefficients in the model are reported in units of eV·nm and eV·nm$^2$, respectively.
\[
H(\bm{k}) = 
\begin{pmatrix}
E_1(\bm{k}) & 0 & 0 & 0 \\
0 & E_1(\bm{k}) & 0 & 0 \\
0 & 0 & E_2(\bm{k}) & 0 \\
0 & 0 & 0 & E_2(\bm{k})
\end{pmatrix}
\]

where
\[
\begin{aligned}
E_1(\bm{k}) &= -1.42 + 0.103(k_x^2 + k_y^2) , \\
E_2(\bm{k}) &= -0.374 + 0.024(k_x^2 + k_y^2).
\end{aligned}
\]

The effective Hamiltonian represents two decoupled electronic bands with parabolic dispersion. It reproduces the DFT band structure accurately within the momentum range $-0.2~\text{\AA}^{-1} \leq k \leq 0.2~\text{\AA}^{-1}$, which is sufficient for reliably determining the Landau level (LL) spectrum of the monolayer. 
The coefficients of the second-order terms $k_x^2$ and $k_y^2$ gives a measure of the effective mass of the respective bands. Based on these coefficients, the extracted effective masses are $1.59m_0$ for the $\Gamma_9$ band and $0.37m_0$ for the $\Gamma_8$ band, where $m_0$ is the free electron mass. This indicates that the $\Gamma_9$ band is significantly heavier than a free electron, whereas the $\Gamma_8$ band is considerably lighter.

\begin{figure}[!!t]
\centering
\includegraphics [width=8cm]{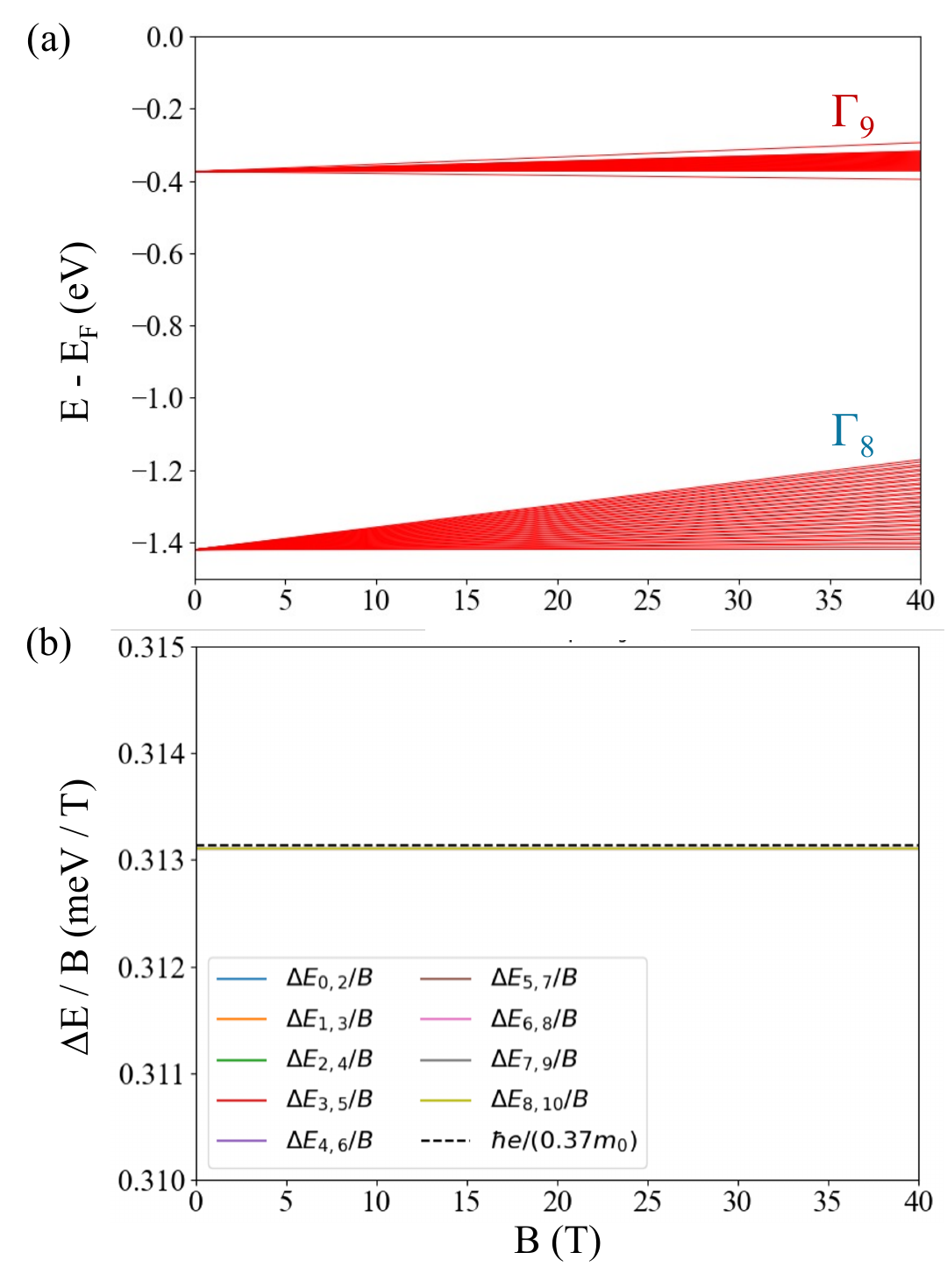}
\caption{(a) Landau level spectrum computed for the Ca$_2$N monolayer with the magnetic field $B$ varying from 0 to 40T. (b) Energy splitting between successive Landau levels as a function of $B$. The dashed line denotes the theoretical value $\hbar e / m^*$, with $m^* = 0.37\,m_0$.}
\label{fig:landau_level}
\end{figure}

To obtain the LL spectrum, we employ the standard quantization procedure as used in previous studies~\cite{kaplan_2024_arxiv}, applying a perpendicular magnetic field $\mathbf{B} = B \hat{z}$ via the Landau gauge $\mathbf{A} = (0, Bx, 0)$. This modifies the momentum operator as $\mathbf{p} \rightarrow \mathbf{p} + e\mathbf{A}$. This choice ensures that the Hamiltonian commutes with $p_y$, i.e., $[H, p_y] = 0$, allowing us to replace $p_y$ by its eigenvalue $\hbar k_y$. This introduces the guiding center coordinate $x_0 = \frac{\hbar k_y}{eB}$, leading to a shifted coordinate $x' = x + x_0$.

To express the Hamiltonian in terms of harmonic oscillator ladder operators, we define

$$
x' = \sqrt{\frac{l_B^2}{2}} (a + a^\dagger), \quad p_{x'} = \frac{i\hbar}{\sqrt{2} l_B} (a^\dagger - a),
$$
where the magnetic length is $l_B = \sqrt{\frac{\hbar}{eB}}$. These operators satisfy the commutation relations $[a, a^\dagger] = 1$ and $[x', p_{x'}] = i \hbar$. To incorporate spin effects, we include a Zeeman coupling term for each band,

$$
H_Z = \frac{1}{2} g \mu_B B \sigma_z,
$$
where $g$ is the Landé $g$-factor, $\mu_B$ the Bohr magneton, and $\sigma_z$ the Pauli spin matrix. Diagonalizing the full Hamiltonian yields the Landau level energies, which are plotted as a function of magnetic field in Figure~\ref{fig:landau_level}(a).

The LLs corresponding to both the $\Gamma_8$ and $\Gamma_9$ states exhibit a linear dependence on the applied magnetic field. This linear dispersion, together with the absence of any crossings between successive LLs, indicates that SOC effects are negligible in the electride states. It also suggests that exchange interactions have minimal influence on the IAEs. Interestingly, the LLs associated with the $\Gamma_8$ band show greater dispersion with increasing magnetic field strength compared to those of the $\Gamma_9$ band. This difference arises from the smaller effective mass of the $\Gamma_8$ band relative to the heavier $\Gamma_9$ band. Since the energy of the $n$th LL is given by $E_n = \left(n + \frac{1}{2}\right)\hbar \omega_c$, where $\omega_c = \frac{eB}{m^*}$ is the cyclotron frequency, the energy dispersion with magnetic field $B$ is therefore inversely proportional to the effective mass $m^*$.

\begin{figure}[!!b]
\centering
\includegraphics [width=8.2cm]{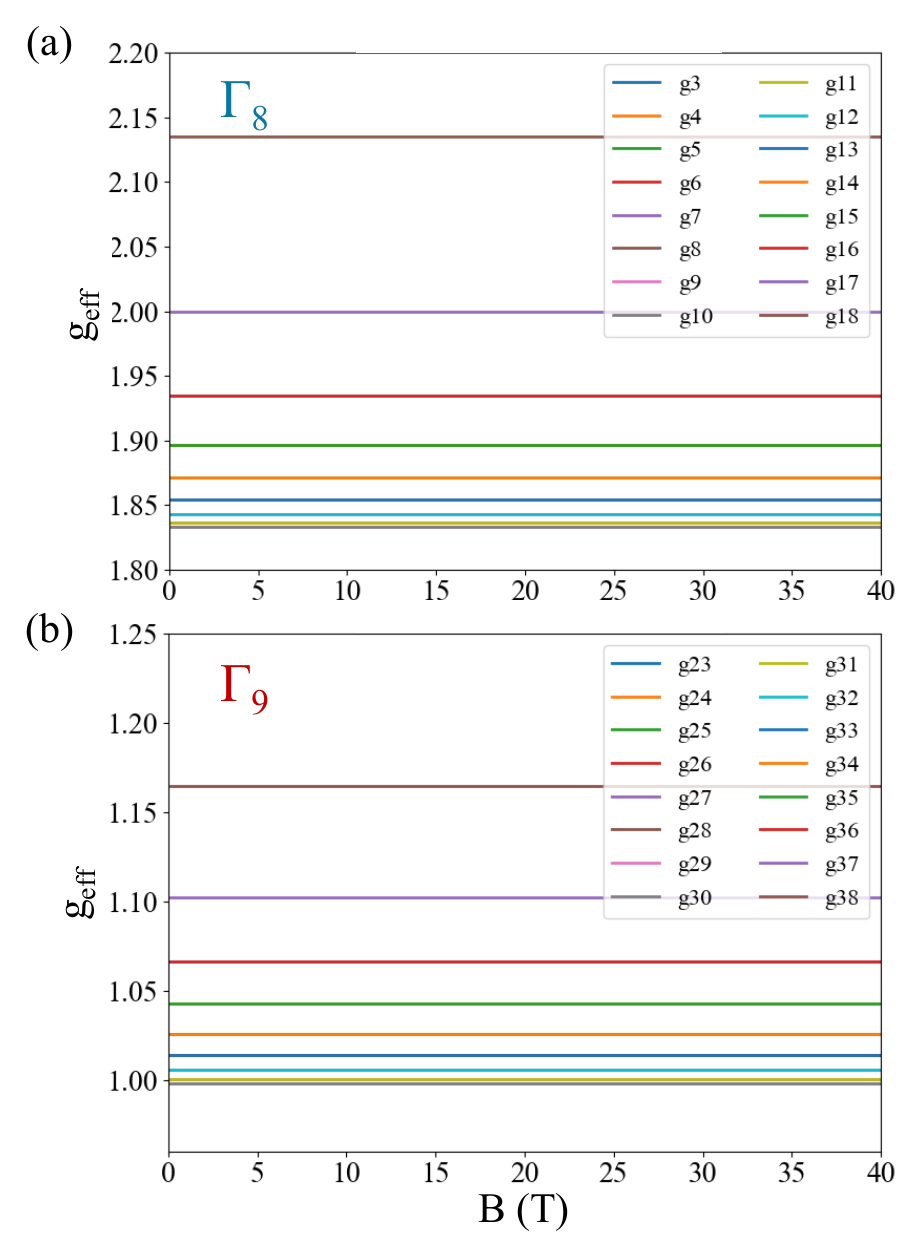}
\caption{Effective g-factor as a function of magnetic field $B$ for (a) the $\Gamma_8$ band and (b) the $\Gamma_9$ band.}
\label{fig:g_factor}
\end{figure}

We then analyzed the energy spacing between consecutive LLs denoted as $\Delta E$ as a function of the applied magnetic field. Figure~\ref{fig:landau_level}(b) shows the variation of $\Delta E/B$ for a representative set of LLs associated with the $\Gamma_8$ state, plotted as a function of the magnetic field. As expected, the energy spacing scales linearly with magnetic field strength, resulting in a constant $\Delta E / B$ ratio. This behaviour follows the relation $\Delta E / B = \hbar\omega_c/B = \hbar e/m^*$. The dashed line in Figure~\ref{fig:landau_level}(b) represents the theoretical value of $\Delta E / B$ ratio calculated using an effective mass of 0.37$m_0$ for the $\Gamma_8$ state. This behavior is consistent with that of a free electron gas.

Finally, we compute the effective $g$-factor for these Landau levels using the relation
$$g_{\mathrm{eff}} = \frac{E_{n\uparrow} - E_{n\downarrow}}{\mu_B B}$$
where $E_{n\uparrow}$ and $E_{n\downarrow}$ are the spin-resolved Landau level energies. Figure~\ref{fig:g_factor} shows the variation of $g_{\mathrm{eff}}$ as a function of magnetic field for the $\Gamma_8$ and $\Gamma_9$ states. Since the electride bands are largely unaffected by SOC, only pure Zeeman splitting is observed in the LL spectrum. Consequently, the effective $g$-factor remains nearly constant with increasing field. For the $\Gamma_8$ band, $g_{\mathrm{eff}}$ lies in the range of 1.8–2.15, close to the free-electron limit. In contrast, the $\Gamma_9$ band exhibits a suppressed $g_{\mathrm{eff}}$ in the range of 1.0–1.2. This decrease is likely due to the reduced band dispersion and higher carrier density associated with the $\Gamma_9$ band, which can suppress the effective $g$-factor, consistent with the findings of Janak \textit{et al.}~\cite{Janak_PhysRev_1969}.

\section{Conclusion}

In this work, we have explored the response of interstitial anionic electrons to high magnetic fields. We computed the Landau level spectrum of Ca$_2$N monolayers corresponding to the electride states that primarily form the Fermi surface. The LLs exhibit a linear dispersion with the applied magnetic field, similar to that of a free two-dimensional electron gas. However, the effective electron mass deviates significantly from the free electron mass. We obtain effective $g$-factors in the range of 1–2, which are close to the free electron limit. Moreover, the invariance of the energy dispersion of the electride bands upon changing the XC functional from LDA to PBEsol indicates the absence of significant local exchange and Coulomb correlation effects in the electride states.

Altogether, these findings suggest that the IAEs in 2D electrides tend to behave as a nearly free electron gas. While such behavior is expected for highly delocalized 2D electride states, investigating the effects of magnetic fields on lower-dimensional electrides (1D and 0D) and exploring how confinement influences the properties of IAEs would be an interesting direction for future research.

\section*{Acknowledgements}
A.B.~and S.S.~acknowledge support from the U.S.~Department of Energy, Office of Science, Office of Fusion Energy Sciences, Quantum Information Science program under Award No.~DE-SC-0020340.
D.K.~is supported by an Abrahams postdoctoral fellowship of the Center for Materials Theory, Rutgers University and the Zuckerman STEM fellowship. D.K.~thanks the hospitality of the Aspen Center for Physics, which is supported by National Science Foundation grant PHY-2210452. D.K.~also acknowledges partial support by grant NSF PHY-2309135 to the Kavli Institute for Theoretical Physics (KITP) where part of this work was finalized.
Authors thank the Pittsburgh Supercomputer Center (Bridges2) supported by the Advanced Cyberinfrastructure Coordination Ecosystem: Services \& Support (ACCESS) program, which is supported by National Science Foundation grants \#2138259, \#2138286, \#2138307, \#2137603, and \#2138296.

\bibliography{bibfile}
\end{document}